\documentclass[preprint,aps,prb,showpacs]{revtex4}

\begin{document}
\title{Superconducting transition of a two-dimensional Josephson junction array in weak magnetic fields}
\author{In-Cheol Baek}
\author{Young-Je Yun}
\author{Jeong-Il Lee}
\author{Mu-Yong Choi}
\affiliation{BK21 Physics Division and Institute of Basic Science,
Sungkyunkwan University, Suwon 440-746, Korea}

\begin{abstract}
The superconducting transition of a two-dimensional (2D) Josephson junction array exposed to weak magnetic fields
has been studied experimentally. Resistance measurements reveal a superconducting-resistive phase boundary in
serious disagreement with the theoretical and numerical expectations. Critical scaling analyses of the $IV$
characteristics indicate contrary to the expectations that the superconducting-to-resistive transition in weak
magnetic fields is associated with a melting transition of magnetic-field-induced vortices directly from a
pinned-solid phase to a liquid phase. The expected depinning transition of vortices from a pinned-solid phase to
an intermediate floating-solid phase was not observed. We discuss effects of the disorder-induced random pinning
potential on phase transitions of vortices in a 2D Josephson junction array.
\end{abstract}


\maketitle

When a two-dimensional (2D) Josephson junction array (JJA) is immersed in a perpendicular magnetic field, a
finite density of vortices are induced in the array. The competition between the repulsive vortex-vortex
interaction and the attractive periodic pinning potential due to the discreteness of the array results in
interesting forms of superconducting-resistive phase boundaries as functions of a magnetic field \cite{R1}. The
superconducting-to-resistive transition in strong magnetic fields or at high vortex densities has been discussed
in terms of the melting of a vortex solid pinned to an underlying lattice. \cite{R2,R3,R4,R5} On the other hand,
the superconducting-to-resistive transition in weak magnetic fields or at low vortex densities was found in
simulation studies to be associated with the depinning transition of a pinned vortex solid, the appearance of
which was predicted by Nelson and Halperin \cite{R6} in late 70s. Monte Carlo (MC) simulations of the lattice
Coulomb gas model for vortices \cite{R7} showed that for a vortex density, the number of flux quanta per
plaquette, $f < 1/25$ the pinned vortex solid transforms into a vortex liquid through two successive phase
transitions: a depinning transition from a pinned solid to a floating solid at a lower temperature $T_p$,
approximately linear in $f$, and a melting transition from a floating solid to a liquid at a higher temperature
$T_m \sim$ 0.045$J/k_B$, approximately independent of $f$, in which $J$ is the Josephson coupling energy per
junction. For $f \gtrsim  1 / 25$, both transitions were found to appear at the same temperature increasing with
$f$. These results have been confirmed semiquantitatively by different numerical or analytical studies of related
models such as the $XY$ model \cite{R8}, the resistively-shunted-junction model \cite{R9}, and the continuum
model of a Coulomb gas in a periodic potential \cite{R10}. While the existence of an intermediate hexatic phase
just above the melting point has been also predicted by theory \cite{R6,R11}, no evidence was found in the
numerical studies.

Despite the agreement among the theoretical and numerical efforts, the experimental confirmation of the
appearance of a depinning transition in the weak field limit has yet to be done. We, therefore, investigated
experimentally the superconducting transition of a JJA in weak magnetic fields. The measurements of the
resistance and the current-voltage ($IV$) characteristics of a square array represent a phase diagram in serious
disagreement with the theoretical and numerical expectations. The expected depinning transition was not observed.
We discuss effects of the disorder-induced random pinning potential on phase transitions of vortices in a 2D JJA.

The experiments were performed on a square array of 400$\times$600 Nb/Cu/Nb Josephson junctions. Cross-shaped
0.2-$\mu$m-thick Nb islands, fabricated by applying photolithography and reactive-ion etching to the Nb/Cu double
layer, were disposed on a 0.3-$\mu$m-thick Cu film periodically with a lattice constant of 13.7 $\mu$m, a
junction width of 4 $\mu$m, and a junction separation between adjacent sites of 1.5 $\mu$m. The photograph of
Fig.\ \ref{fig1}(a) shows the structure of the JJA sample used in this work. The variation of the junction
separation in the sample was less than 0.1 $\mu$m. The standard four-probe technique was adopted for the
measurements of the resistance and the $IV$ characteristics. The current and voltage leads were attached to the
Cu layer. The current leads were connected 3 mm away from the ends of the array to improve the uniformity of the
injected current. Voltage leads were placed 100 junctions away from the end edges of the array so that there
exist 400$\times$400 junctions between the voltage leads. The sample voltage was measured by a
transformer-coupled lock-in amplifier with a square-wave current at 23 Hz. The single-junction critical current
$i_c$ at low temperatures can be obtained directly from the $IV$ curve. At low temperatures, the slope $dV/dI$ of
the curve has its maximum at a current equal to $i_c$ multiplied by the number of Nb islands perpendicular to the
current direction. The $i_c$ and the junction coupling strength $J(=\hbar i_{c}/2e)$ at high temperatures were
determined by extrapolating the $i_c$ vs $T$ data at low temperatures by the use of the de Gennes formula for a
proximity-coupled junction in the dirty limit, $i_{c}(T) = i_{c}(0)(1-T/T_{co})^{2}\exp(-\alpha T^{1/2})$, where
$T_{co}$ is the BCS transition temperature. We have also tried another formula developed by Zaikin and Zharkov
\cite{R12} and found both formulas generate practically the same $i_c$'s for our sample at the temperatures of
interest. Representing the superconducting transition temperatures of the sample in units of $J/k_{B}$ by using
the $i_{c}$'s determined with the de Gennes formula, they are 0.85 for $f=$ 0 and 0.42 for $f=$ 1/2. These values
are lower by 5-7\% than the superconducting transition temperature ($T_{c}$'s) found from many numerical studies,
which are 0.89 for $f=$ 0 and 0.45 for $f=$ 1/2 \cite{R13}. This demonstrates that $i_{c} (T)$ of the sample
determined with the formulas is accurate within 10\% of error. During the measurements, the temperature was
controlled by a Lake Shore 340 temperature controller with fluctuations less than 1 mK. A solenoid generated
external magnetic fields in the sample space where ambient magnetic fields were screened out by $\mu$-metal. The
magnetic field or the vortex density was adjusted from the magnetoresistance measurements of the sample with
50-$\mu$A excitation current. The presence of distinct resistance minima at fractional $f$'s in the magnetic
field vs resistance curve made precise adjustment of the vortex density possible.

Figure\ \ref{fig1}(c) shows the sample resistance with 30-$\mu$A excitation current as a function of $f$ at $T=$
3.59 K. The appearance of many pronounced higher-order dips indicates the good uniformity of the magnetic field
over the sample. The magnetoresistance curve is impressively similar to the mean-field superconducting-resistive
phase boundary numerically predicted in Ref. 1. The curve exposes, as numerically predicted, that the mean-field
superconducting transition temperature $T_{MF}$ is a decreasing function of $f$ in the low $f$ region. As stated
above, $T_{c}$ as an increasing function of $f$ was observed in the same region in the MC simulations. Even
though a mean-field phase diagram does not always resemble the real one, the $f$-dependence of $T_{MF}$ of our
sample raises a question about validity of the phase diagram from the MC simulations. In order to examine the
$f$-dependence of $T_{c}$ of a JJA in weak magnetic fields, we measured the sample resistance $R$ as functions of
the temperature $T$ for  six different vortex densities $f = 1 / 50,~ 1 / 36,~ 1 / 25,~ 1 / 16,~ 1 / 10$, and $1
/ 8$. Figure\ \ref{fig2} shows some of the $T$ vs $R$ traces. The superconducting transition temperature $T_c$
determined from the $T$ vs $R$ curves is plotted against $f$ in Fig.\ \ref{fig3}. The superconducting-resistive
phase boundary differs from the theoretical and numerical expectations summarized above. $T_c$ decreases with $f$
increased, contrary to the numerical observations. The $T_c$ at $f = 1/50$ is $\sim$0.32 in units of $J/k_B$,
which is $\sim$10 times the numerical value of $T_p$ and $\sim$7 times that of $T_m$. The $T_c$'s at $f = 1/50,
1/36$, and 1/25 exceed even the melting temperatures of densely populated vortices for $f =$ 2/5, 1/3, and 1/5
\cite{R4,R5,R14}. Such high $T_c$'s appear hardly reconcilable with the proposed depinning transition, but
possibly with the melting transition of a pinned vortex solid as for dense vortex systems.

The gradual drop of the resistance of the sample upon approaching the transition indicates that the observed
superconducting transitions are all continuous phase transitions. For a continuous superconducting transition,
the critical scaling analysis of the $IV$ characteristics may provide valuable information about the nature of
the superconducting transition. Figure\ \ref{fig4} displays the $IV$ characteristics of the sample for $f =$
1/50, 1/36, and 1/16. The ${\log}I$ vs ${\log}V$ isotherms were obtained by averaging 15-240 measurements for
each value of current. The derivative $d({\log}V)/d({\log}I)$, that is, the slope of the $IV$ curves of Fig.\
\ref{fig4} as functions of $I$ are in Fig.\ \ref{fig5}. The data at high currents are due to the single-junction
effect and thus should be left out of consideration. The derivative plots show that for all the $f$'s studied,
the $IV$ curves have negative curvatures at low temperatures. At high temperatures, they have positive curvatures.
The high-temperature and low-temperature curves are separated by a straight line satisfying the power-law $IV$
relation. Thermally activated vortex motion would give rise to Ohmic $IV$ characteristics in the low current
limit and positive curvature in $IV$ curves at intermediate currents. Negative curvature in the low-$T$ curves
implies vanishing resistance in the low current limit, indicating a superconducting state as the low-temperature
state. The $IV$ characteristics thus indicate that for the $f$'s, the JJA undergoes a continuous superconducting
transition at the temperature where a straight $IV$ curve appears. A detailed scaling theory \cite{R15} suggests
that for a continuous superconducting transition, the $IV$ characteristics in 2D should scale as
$V/I|T-T_{c}|^{z\nu} = {\cal E}_{\pm}(I/T|T-T_{c}|^{\nu})$, where $\nu$ and $z$ are critical exponents and
$\cal{E}_{\pm}$ the scaling functions above and below $T_c$. This scaling form becomes a simple power-law $IV$
relation, $V \sim I^{z+1}$, at $T = T_c$. Thus, one may find $T_{c}$ and the dynamic critical exponent $z$
directly from the straight ${\log}I$ vs ${\log}V$ isotherm. The correlation-length exponent $\nu$ can be obtained
from the scaling analysis of the $IV$ data. Although the scaling analysis of $IV$ data is an effective tool for
studying the critical behaviors of superconductors, special care should be taken not to extract any false
information from the analysis. It has been recently shown for cuprate superconductors that the good data collapse
can be achieved for a wide range of $T_{c}$ and critical exponents and sometimes does not even prove the
existence of a phase transition. \cite{R16} It has been also shown that both the current noise \cite{R17} and the
finite-size effect \cite{R18} may create Ohmic behavior at low currents even below the superconducting transition
and lead to an underestimate of $T_{c}$ and incorrect $\nu$ and $z$. The $IV$ data in Fig.\ \ref{fig4}, however,
do not suffer such complications. We avoided the current-noise problem by employing the phase-sensitive
signal-detection method using a low-frequency square-wave current as described above. The effect of
finite-size-induced free vortices was also found insignificant for JJA samples exposed to a magnetic field.
\cite{R5} The $IV$ curves in Fig.\ \ref{fig4} prove that the large flexibility in determining the critical
exponents and temperature from $IV$ data is not the case for the array sample. Unlike the $IV$ data of cuprate
superconductors, the $IV$ curves of the array exhibit evident concavities below the transition. One can determine
$T_{c}$ with uncertainty $\Delta T_{c} \approx \pm 0.1$ K from the curves. The $T_{c}$'s determined from the
straight $IV$ curves agree within the uncertainties with those from the $T$ vs $R$ data. In the process of the
scaling analysis, the distinct concavities of the $IV$ curves further reduce the arbitrariness in determining
$T_{c}$ to $\pm$0.5-0.7 K. The $IV$ data scaled on the basis of the scaling form with $T_c$ and $z$ derived from
the straight $IV$ curves are shown in Fig.\ \ref{fig6}. Each plot contains $IV$ curves at 18-19 different
temperatures. The scaling plots confirm that the JJA at the $f$'s studied undergoes a continuous superconducting
phase transition at the temperature where a straight $IV$ curve appears. The insets of Fig.\ \ref{fig6} exhibit
the values of $T_c$, $\nu$, and $z$ used to scale the data. The $T_c$'s derived from the scaling analyses are in
accordance within experimental errors with those from the resistance measurements. For all three $f$'s, the
dynamic critical exponent $z$ is less than 1 and the correlation-length exponent $\nu$ is much larger than the 2D
Ising value ($\nu_{I}=1$). The low-temperature $IV$ curves can be fitted into the form $V \sim I
\exp[-(I_{T}/I)^{\mu}]$ with $\mu =$ 0.7-1.1. The scaling behaviors of the $IV$ characteristics are quite similar
to those found for $f =$ 2/5, 1/3, and 1/5 \cite{R5}, for which the superconducting transition is understood in
terms of melting of a pinned vortex solid driven by domain-wall excitations \cite{R3,R4,R5}. The large
low-current voltage signals near the transition also appear to be compatible with the melting transition at
$T=T_{c}$. A floating solid is expected to be much less mobile near the transition and thus generate much lower
voltage signal than melted vortices. Taking into consideration of the low vortex densities of our systems, the
low-current voltage signals near the transition are as large as those of melted vortices for $f=$ 2/5, 1/3, and
1/5 reported in Ref. 5. The large voltage signals above the transition, the scaling behaviors of the $IV$ data,
and the high $T_{c}$'s, all indicate that the observed superconducting transitions in the weak magnetic fields
are associated with the melting transition of a pinned vortex-solid to a vortex liquid without passing through an
intermediate floating solid phase.

The high melting transition temperatures and the absence of a depinning transition are in serious disagreement
with the theoretical and numerical expectations. The disagreement in weak magnetic fields contrasts with the
agreement in strong magnetic fields. For $f=$ 2/5, the same sample exhibited a superconducting transition at
about the same temperature as observed in simulations. It has been recently found in numerical studies of the 2D
Coulomb gas \cite{R19} that the floating vortex-solid phase may exist only in small arrays. This may explain the
absence of a floating solid phase in our large array sample, but not the high melting temperatures. We therefore
consider the disorder-induced random pinning potential as a possible cause for the disagreement in weak magnetic
fields. The numerically obtained phase diagram \cite{R7,R10} intimates that the depinning and melting transitions
may appear at the same temperature when the depinning temperature $T_p$ exceeds the melting temperature $T_m$. A
higher $T_p$ can be achieved with stronger pinning of vortices to the underlying lattice. In the process of
fabricating a JJA sample, random variation of the junction coupling strength is inevitably introduced. The
inevitable variation of junction coupling strength in a real array may provide additional pinning and an enhanced
$T_c$ for vortices dislocated from the periodic disposition. Such an effect of random bond disorder on $T_c$
should be much weaker in strong magnetic fields, consistent with the observations, because when densely populated
vortices are dislocated from the periodic disposition, the increment of the repulsive vortex-vortex interaction
energy effectively countervails the contribution of the random pinning potential. The variation of the coupling
strength in our sample estimated from the $I$ vs $dV/dI$ curve in Fig.\ \ref{fig1}(b) is $\lesssim \pm$15\%. Yet
it is not quite certain whether such an amount of variation of the coupling strength can raise $T_c$ so much. In
order to ascertain whether extra random pinning can raise $T_c$ so effectively in weak magnetic fields, we
measured resistively the $T_c$ of a system with artificially introduced random pinning. Figure 7 shows the $T_c$
as a function of $f$ for a square JJA in which 14\% of superconducting islands are randomly diluted. This
site-diluted sample also contains the inevitable bond disorder, that is, the variation of junction coupling
strength as much as the foregoing sample. We find in the figure that the random site disorder affects the $T_c$
of a JJA as presumed above. The $T_c$ in units of $J/k_{B}$ for $f=$ 1/50-1/8 was elevated by the addition of
site-disorder by 47-100\% above that of the foregoing sample. For $f=$ 2/5, the change of $T_c$ was found less
than 5\%. The superconducting behavior of the site-diluted sample proves that the random pinning potential
induced by disorder is quite effective in stabilizing the superconducting vortex-solid phase in weak magnetic
fields.

In summary, the measurements of the resistance and the $IV$ characteristics disclosed a phase diagram of a JJA
exposed to weak magnetic fields in sharp contrast to the theoretical and numerical expectations. Vortices in a
square array at $f=$ 1/50-1/8 were found to undergo a melting transition directly from a pinned-solid phase to a
liquid phase at $T_{c} \sim$ 0.15-0.32 $J/k_B$, $\sim$3-7 times the predicted melting temperature. The expected
depinning transition was not observed. The experiments on the site-diluted array reveal that the disorder-induced
random pinning can stabilize the superconducting pinned-solid phase in weak magnetic fields even at 10 times the
melting temperature of a disorder-free system and is probably responsible for the experimental findings contrary
to the theoretical expectations.

This work was supported by the BK 21 program of the Ministry of Education.

\newpage

\begin{figure}[h!]
\caption{(a) Photograph of the square Josephson junction array used in the experiment. (b) $I$ vs $dV/dI$ at $T =$ 3.10 K. (c) Sample resistance with 30-$\mu$A excitation current as a function of the vortex density $f$ at T = 3.59 K.}
\label{fig1}
\end{figure}

\begin{figure}[h!]
\caption{Temperature dependences of the resistance at $f =$
(a) 1/50,
(b) 1/36, and
(c) 1/16.
The arrows denote where the superconducting transition occurs. The resistance was measured with 3 $\mu$A current injected into the sample.}
\label{fig2}
\end{figure}

\begin{figure}[h!]
\caption{Vortex-density dependence of $T_c$ determined from the resistance measurements. Note that $T_c$ is in units of $J/k_B$.}
\label{fig3}
\end{figure}

\begin{figure}[h!]
\caption{The IV characteristics for
(a) $f = {1 / 50}$ at $T$ = 3.600-4.500 K,
(b) $f = {1 / 36}$ at $T$ = 3.600-4.500 K, and
(c) $f = {1 / 16}$ at $T$ = 3.500-4.400 K.
The dashed lines are drawn to show the power law ($V \sim I^{z+1}$ ) behavior at the superconducting temperature.}
\label{fig4}
\end{figure}

\begin{figure}[h!]
\caption{Current dependences of the slope of the $IV$ curves in Fig.\ \ref{fig4}. Excessively fluctuating data at low currents are not shown.}
\label{fig5}
\end{figure}

\begin{figure}[h!]
\caption{Scaling plots of the $IV$ curves. Each plot contains $IV$ curves at 18-19 different temperatures. The
insets show the values of $T_c$, z, and $\nu$ used to scale the data.} \label{fig6}
\end{figure}

\begin{figure}[h!]
\caption{$T_c$ of the 14\% site-diluted sample as a function of vortex density.}
\label{fig7}
\end{figure}

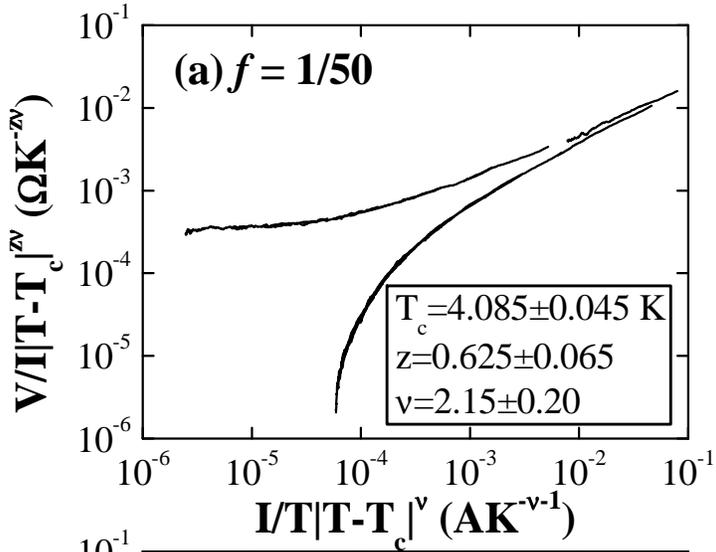
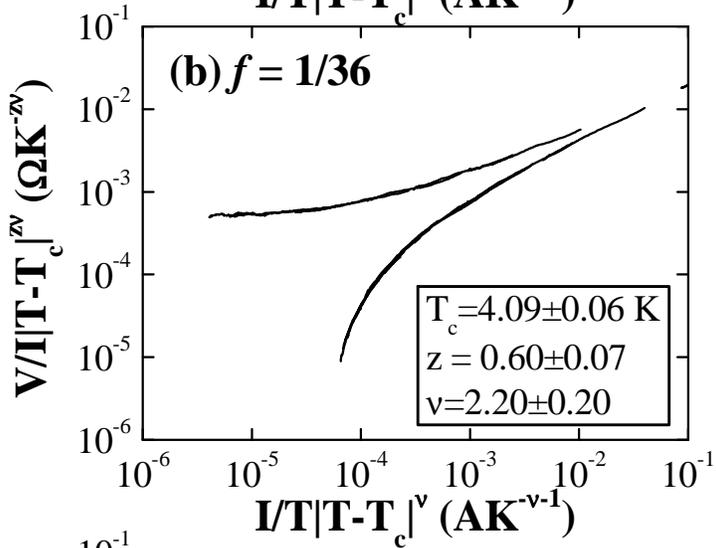
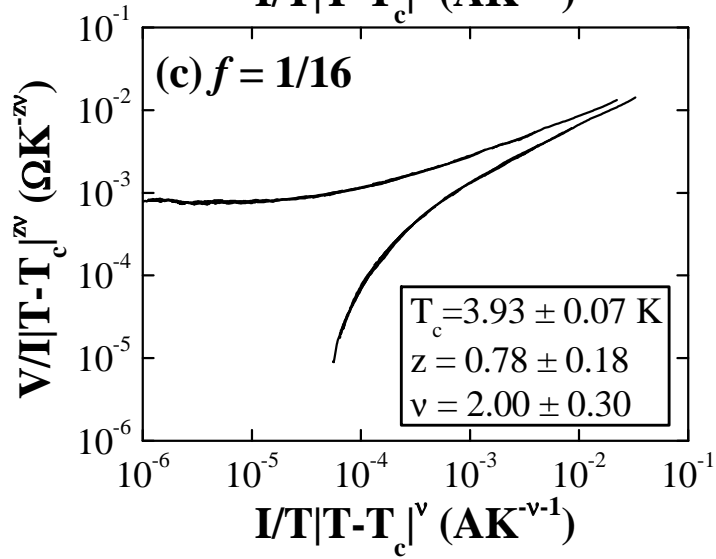
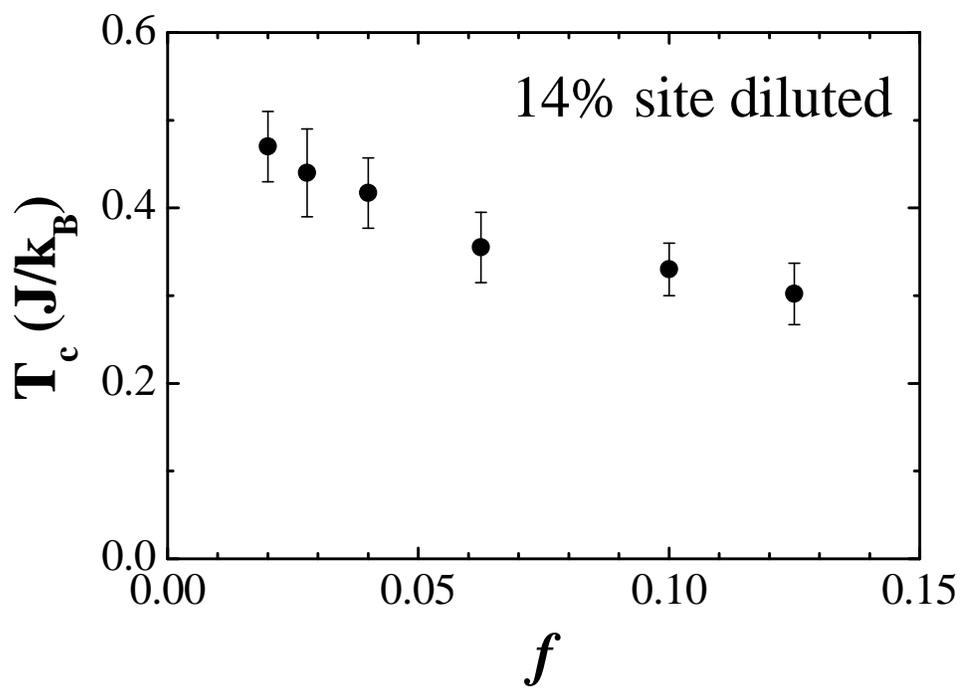
\end{document}